\documentclass[12pt,preprint]{aastex}

\shorttitle{The eclipsing binary V32 in NGC~6397}
\shortauthors{Kaluzny et al.}

\begin{document}

\title{The Clusters AgeS Experiment (CASE).~III. \\
Analysis of the Eccentric Eclipsing
Binary V32 in the Globular Cluster 
NGC~6397\footnote{This paper utilizes data 
obtained with the 6.5-meter Magellan Telescopes located at Las Campanas 
Observatory, Chile.}
}

\author{ 
J.~Kaluzny\altaffilmark{2},
I.~B.~Thompson\altaffilmark{3},
S.~M.~Rucinski\altaffilmark{4}, 
W.~Krzeminski\altaffilmark{5},
}

\altaffiltext{2}{Copernicus Astronomical Center, Bartycka 18,
00-716 Warsaw, Poland; jka@camk.edu.pl}

\altaffiltext{3}{Carnegie Observatories, 813 Santa Barbara St.,
Pasadena, CA 91101-1292; ian@ociw.edu}

\altaffiltext{4}{David Dunlap Observatory, Department of Astronomy and 
Astrophysics, University of Toronto, P.O. Box 360, Richmond Hill, 
ON L4C 4Y6, Canada; rucinski@astro.utoronto.ca}

\altaffiltext{5}{Las Campanas Observatory, Casilla 601, La Serena, 
Chile; wojtek@lco.cl}

\begin{abstract}
We present spectroscopic and photometric observations of the eclipsing binary
V32 located in the central field of the globular cluster NGC~6397.
The variable is a single-line  spectroscopic binary with an orbital
period of 9.8783~d and a large eccentricity of $e=0.32$.
Its systemic velocity
($\gamma=20.7$ km~s$^{-1}$) and metallicity ([Fe/H] $\sim$ -1.9) are both 
consistent with cluster membership.   
The primary component of the binary is located at the
top of the main-sequence turn-off on the cluster color-magnitude diagram.
Only a shallow primary eclipse is observed in the light curve.
Based on stellar models for an age of 12~Gyr and the mass-function derived from 
the radial velocity curve, we estimate the masses to be $M_{p}=0.79$~M$_{\odot}$
and $M_{s}=0.23$~M$_{\odot}$. 
The light curve of V32 can be  reproduced by adopting
$R_{p}=1.569$~R$_{\odot}$ and $R_{s}=0.236$~R$_{\odot}$
for the radii and $i=85.44$~deg for the system inclination.
The system geometry precludes observations of the secondary
eclipse. The large eccentricity of the orbit is puzzling
given that for metal poor, halo binaries the transition from circular
to eccentric orbit occurs at an orbital period of about 20~days.
We suppose that the orbit of V32 was modified relatively recently
by dynamical interaction with other cluster star(s). 
An alternative explanation of the observed eccentricity 
calls for  the presence of a third body in the system.
\end{abstract}


\section{INTRODUCTION}
\label{intro}

Little is known about the properties of main-sequence binary stars in 
globular clusters. These systems offer the potential for 
measurements of cluster age and distance independent of cluster
main-sequence fitting \citep[see, for example, ][]{pac,thomp01},
and as  testbeds for theories of the stellar evolution of Population-II
stars.

The eclipsing binary V32-N6397 (hereafter V32) was discovered by 
\citet{kal06} during a survey for variable stars in the central field of
the globular cluster NGC~6397. They observed only one shallow eclipse
event with a depth of about 30 mmag. With $V_{max}=16.13$ the
variable is located at the very top of the cluster main sequence, 
and is a candidate member of NGC~6397.
In this paper we report results of follow-up photometric and
spectroscopic observations of V32.
Section~\ref{obs} describes new photometry and a measurement
of  the spectroscopic orbit of V32. A self-consistent model
of the system is presented in Section~\ref{lc}. 
Finally, Section~\ref{disc} briefly discusses
the implications of the large measured eccentricity of the   
orbit of V32.

\section{SPECTROSCOPIC AND PHOTOMETRIC OBSERVATIONS}
\label{obs}

Spectroscopic observations of V32 stars were carried out
with the MIKE echelle spectrograph \citep{bern03}
on the Magellan~II (Clay) telescope at Las Campanas Observatory.
The data were collected
during several observing runs between 2005 May and 2007 August.
For this analysis we use data obtained with the blue 
channel of MIKE covering the wavelength range 3350~\AA~to 5000~\AA~at 
a resolving power of $\lambda / \Delta \lambda \approx 38,000$.
All of the observations were obtained with a $0.7\times 5.0$ arcsec slit
and with $2\times 2$ pixel binning. At 4380 \AA\ the resolution was
$\simeq 2.7$ pixels with a scale of 0.043~\AA/pixel.
The spectra were first processed using a pipeline developed by Dan
Kelson following the formalism of \citet{kel03} and then analyzed
further using standard tasks in the IRAF/Echelle package\footnote{IRAF
is distributed by the National Optical Astronomy Observatories, which
are operated by the Association of Universities for Research in
Astronomy, Inc., under cooperative agreement with the NSF.}. Each of
the final individual spectra typically consisted of two 300~s exposures
interlaced with an exposure of a thorium-argon lamp. We obtained 16
spectra of V32. 

Velocities were measured with the IRAF FXCOR package. For the template 
we used  a single MIKE 
spectrum of the metal-poor subgiant HD~193901 
\citep[${\rm [Fe/H] = -1.22}$,][]{tom92}
with an adopted radial velocity of $-172$ km~s$^{-1}$. 
The velocity measurements were made over the wavelength range 
4000~\AA~ -- 5000~\AA~with the Balmer lines masked out of the template and 
object spectra. There was no evidence for a second velocity peak in
any of the object spectra and we conclude that the variable is a single-line
spectroscopic binary.
The observations are presented in Table~\ref{tab1} which lists the
heliocentric Julian Date (HJD) at mid-exposure, the velocities of the primary
and errors of  these velocities as returned by the FXCOR routine.

\begin{figure}
\epsscale{1.0}
\plotone{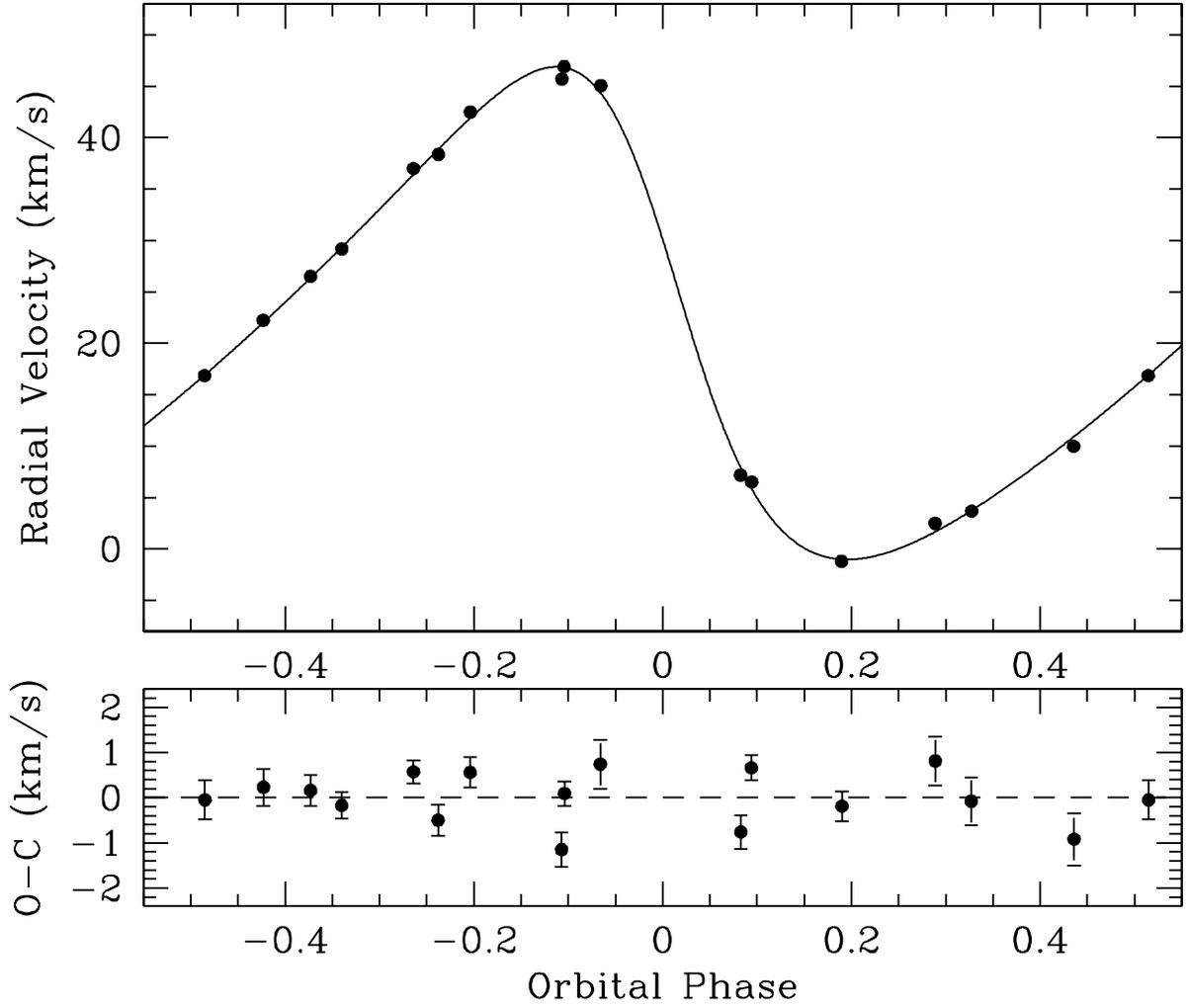}
\caption{
The velocity curve for orbital solution for V32, together with the 
individual observed velocities.  The O-C residuals from the solution
are shown in the lower panel. 
\label{fig1}}
\end{figure}

The measured radial velocities of the primary were fitted with a non-linear
least squares solution of the eccentric orbit using code written by G.\ Torres.
The derived orbital elements are listed in the second column of Table~\ref{tab2}. 
The quantity $T_{1}$ is the moment of periastron passage and the remaining 
quantities have their standard meaning. The solution with 6 free parameters
implies that a superior conjunction occured at $HJD=245 3109.7977 \pm 0.083$.
This is consistent with with the timing of the eclipse observed
on the night of 2004 April 11. Using photometry from \citet{kal06}
supplemented with new observations (see below) we estimate that the 
observed eclipse was centered at $HJD=245 3109.829$.
We then fitted the radial velocity curve by fixing the orbital
period at the value which exactly predicts 
the moment of  superior conjunction. By fitting 5 parameters we obtained
$P=9.8779$~d and $T_{1}=2453899.8248$. This ephemeris predicts
correctly the timing of the eclipse observed in the 2004 season. However,
it leads to unacceptable phasing of some observations from the 2007
season which were obtained shortly before ingress to the primary 
eclipse. The shortest orbital period which gives a symmetric primary 
eclipse  is $P=9.8783$~d. For this period
the observed moment of the observed superior conjunction     
can be reproduced for $T_{1}=2453899.8890$.
The solution with $P$ and $T_{1}$ fixed at these values is listed in 
the third column of Table~\ref{tab2}. The corresponding velocity curve and 
observed velocities are plotted in Fig.~\ref{fig1}.

\begin{figure}
\plotone{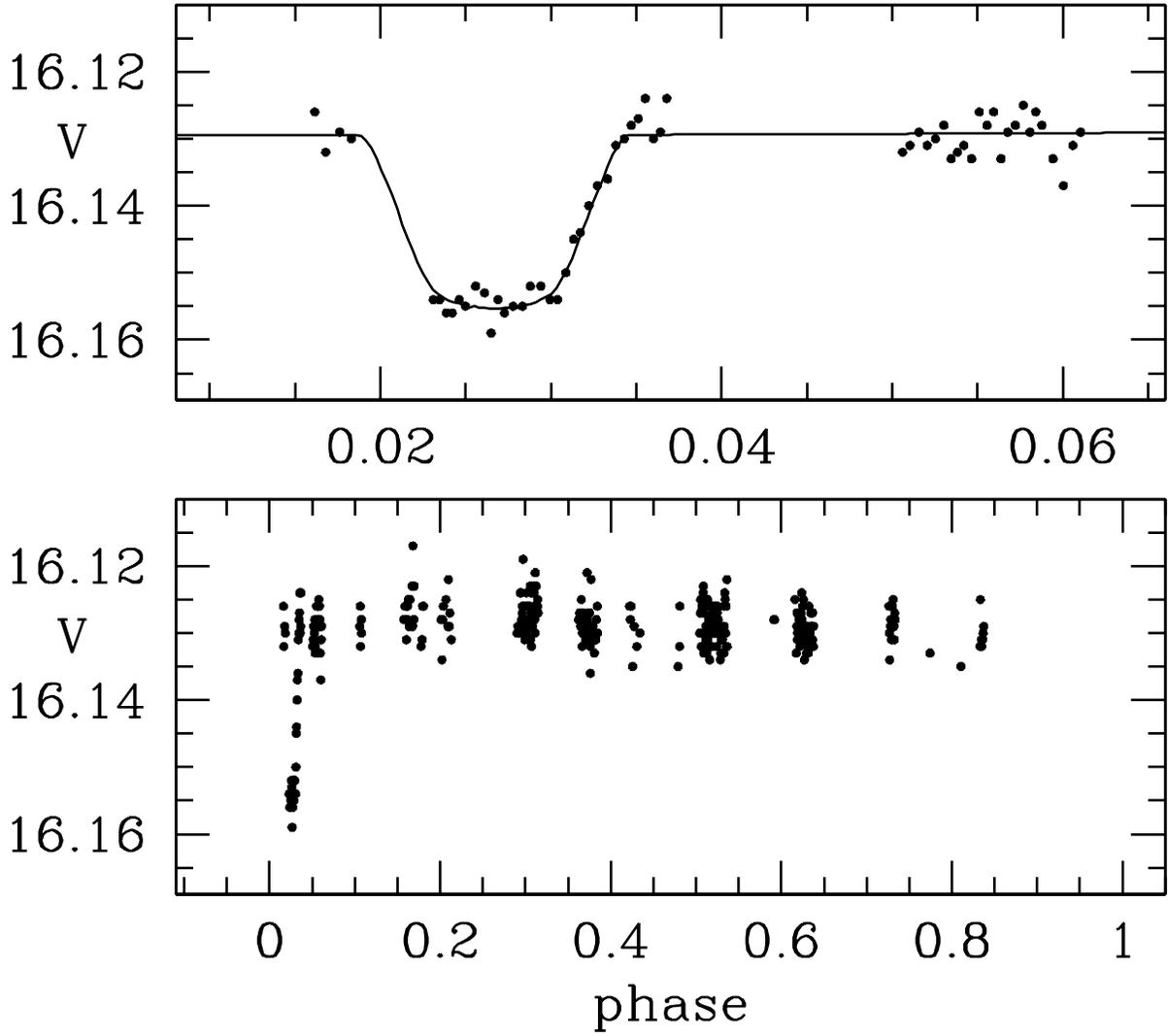}
\caption{
The phased observed $V$ light curve of V32. 
The synthetic curve corresponding to the model described in Sec. 3 
is shown in the upper panel.
\label{fig2}}
\end{figure}

\begin{figure}
\epsscale{1.0}
\plotone{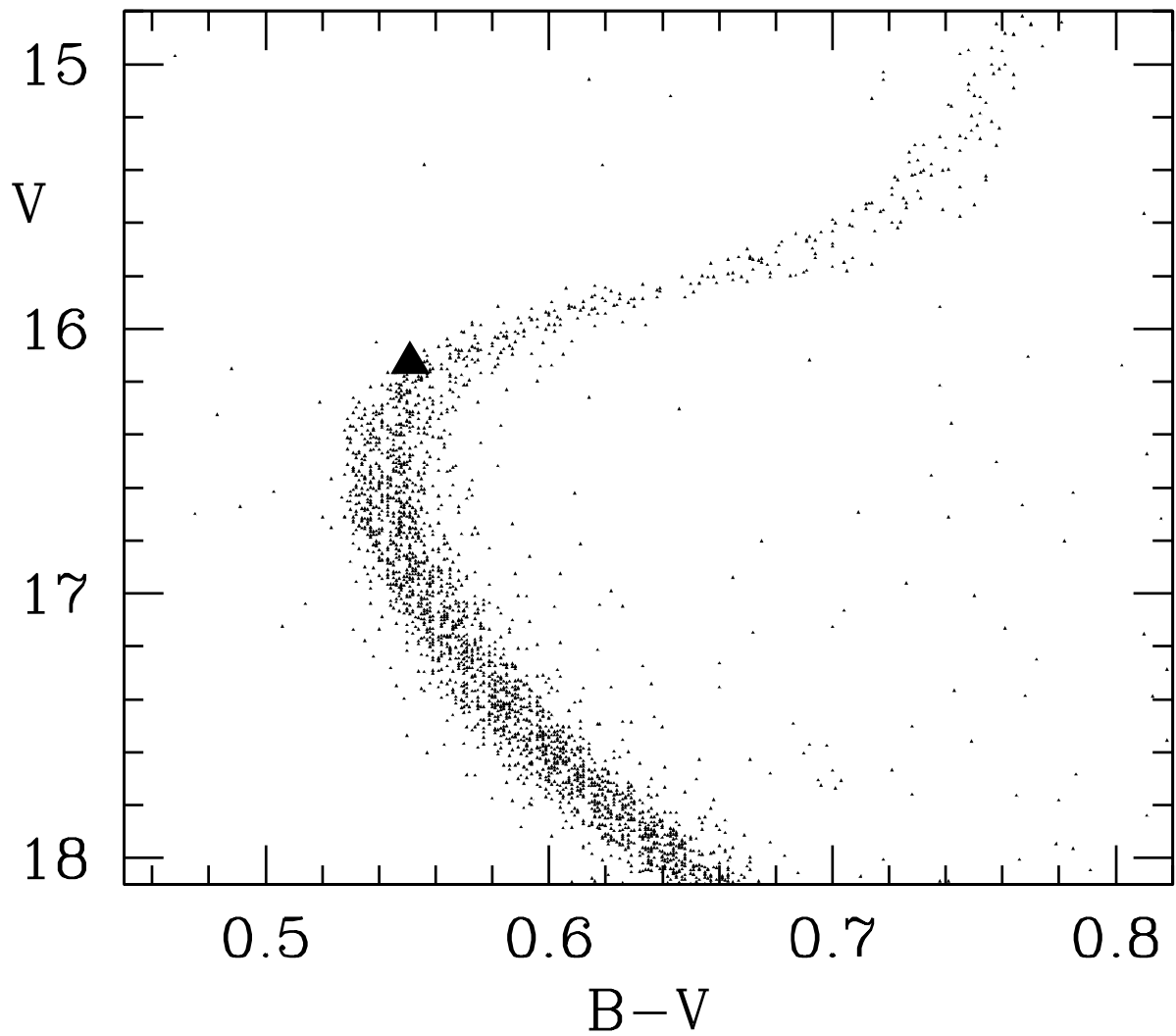}
\caption{
Position of V32 in the $V/B-V$ diagram for NGC~6397. 
\label{fig3}}
\end{figure}

At the beginning of the 2007 observing season we had 
in hand an approximate ephemeris for V32 and further 
photometric observations of a few predicted eclipses of the binary 
were obtained. 
The observations were made with the TEK5 camera on the
2.5m du Pont telescope at Las Campanas Observatory. 
The instrumental setup and reduction methods
were the same as these described in \citet{kal06}.
These observations were mostly hampered by poor weather but
on one of the nights we obtained some useful data covering the 
phases immediately preceding the predicted ingress into primary eclipse.
These observations helped to constrain the spectroscopic solution
as described above.  
The $V$-band light curve based on all available observations of V32 
is shown in Fig.~\ref{fig2} for the region of the light curve
near the observed eclipse.  It is phased with the spectroscopic ephemeris 
listed in the third column of Table~\ref{tab2}. 
The continuous line shows the synthetic
light curve corresponding to the model presented in the next section.
The full $V$-band observation data set is given in Table~\ref{tab3}.
The $V/B-V$ color-magnitude diagram for NGC~6397 based on the du Pont
photometric observations is presented in Fig.~\ref{fig3}.

\section{Properties of V32}
\label{lc}

A direct determination of the absolute parameters of the 
components of V32 is hampered by the lack of velocity 
observations of the secondary. In the following analysis we assume that 
the variable is a  member of the globular cluster NGC~6397.
This conclusion is based on four arguments. First, the systemic 
velocity of V32 ($\gamma=20.65 \pm 0.19$~km~s$^{-1}$) agrees with
the mean velocity of the cluster determined by 
\citet{meylan}, $V_{rad}=18.1\pm 0.1$~km~s$^{-1}$ with  
a central velocity dispersion of 4.5 km~s$^{-1}$. 
Second, on the $V/B-V$ diagram  shown in Fig.~\ref{fig3} 
the binary is located at the top of the cluster turnoff, 
the primary is apparently just beginning to
evolve onto the cluster sub-giant branch.   
Third, the variable is located on the sky only 
51 arcsec from the projected center of  NGC~6397. The cluster
itself has a half-light radius of $r_{h}=140$~arcsec \citep{harris}.
Finally, we have estimated the metallicity of V32. We shifted the 
individual spectra of V32 to zero velocity, and averaged these
shifted spectra to produce a mean spectrum for V32. We measured 
the equivalent widths of 10 Fe~I lines selected from \citet{thomp08}
in this mean spectrum,
and compared these to the equivalent widths measured in a grid of
alpha-enhanced synthetic spectra from \citet{coelho} linearly 
interpolated at $T_{eff}$ = 6400 and $log~g$ = 4.3 for 
metallicities of [Fe/H] = -2.4, -2.0, -1.6, and -1.2. From this 
comparison we obtain
[Fe/H] = -1.9 $\pm$ 0.1 (internal standard deviation) for V32. Recent 
measurements
of the metallicity of NGC~6397 find [Fe/H] = -2.0 \citep{gratton,thevenin}.
While our measurement is crude (a fine analysis of the mean spectrum is
in preparation (McWilliam et al. 2008)), the estimated metallicity of V32 
is consistent with cluster membership.

We adopt a cluster metallicity of
${\rm [Fe/H]=-2.0}$ with an $\alpha$-element 
enhancement of ${\rm [\alpha/H]=+0.34}$.
Recent estimates of the age of NGC~6397 span a range of
$11.47 \pm 0.47$~Gyr \citep{hansen}
to $13.5\pm 1.1$ (\citet{gratton}; 
a determination for models with diffusion). 
We  adopt an age of $12.0 \pm 0.5$~Gyr.
Using the Dartmouth stellar model isochrones of \citet{dotter}
for an age of $12.0 \pm 0.5$~Gyr,
${\rm [Fe/H]=-2.0}$ and ${\rm [\alpha/H]=+0.4}$, the
location of the primary of V32 on the cluster $V/(B-V)$ diagram
implies a primary mass of $M_{p}=0.790\pm 0.009$~M$_{\odot}$. Assuming for
the moment an orbital inclination of  $i=90$~deg, we use the
measured mass function to derive a
secondary mass of $M_{s}=0.232\pm 0.003~M_{\odot} $. The 
\citet{dotter} models then suggest a secondary radius of 
$R_{s}=0.236~R_{\odot}$. 

We fit the light curve of V32 using the PHOEBE code 
\citep{prsa05} which is based on the models of \citet{wd71}
and \citet{w79}. Two  free parameters were fit: the inclination $i$ and the radius of the 
primary. The remaining parameters were fixed as follows:
The orbital elements were adopted from the 3rd column of Table 2.
The effective temperature
of the primary was set at $T_{p}=6490$~K based on an unreddened
$(B-V)_{p}=0.368$
and the empirical calibration of \citet{worthey}.
Following \citet{gratton} we adopted $E(B-V)=0.183$.
The temperature of the secondary was set at 
$T_{s}=3874$ using the Dotter et al. (2007) models.
The solution converged to $i=84.96$~deg and $R_{p}=1.53~R_{\odot}$. 
We then made one more iteration of the above procedure starting 
with $i=84.96$~deg, $M_{s}=0.233~M_{\odot} $
and $R_{s}=0.237~R_{\odot}$. The solution converged to $i=85.03$~deg and
$R_{p}=1.53~R_{\odot}$. The calculated luminosity ratio in the $V$-band
is $L_{p}/L_{s}=664$. In the middle of the primary eclipse the entire
disk of the secondary is projected against disk of the primary -- 
the eclipse is a transit. The secondary (occultation)
eclipse is not observable.
The estimated radius of the primary is consistent with its location
above the main-sequence turnoff on the color-magnitude diagram of 
the cluster. For a mass of $0.790~M_{\odot}$ and an age of 12~Gyr 
the models of  Dotter et al. (2007) predict a radius of $r=1.40~R_{\odot}$.
For the same age the radius reaches $r=1.54~R_{\odot}$ for a 
mass of $0.793~M_{\odot}$. Given the approximate nature of our analysis
we consider the derived solution to be self-consistent.

Clearly the limited data make a direct and accurate determination
of the absolute parameters for components of V32 impossible. The light curve lacks 
a secondary eclipse and the photometric coverage
of the primary eclipse is not complete. However, the deduced parameters 
of V32 reproduce well its observed radial velocity and light curves, and are 
self-consistent with evolutionary models 
of low mass stars.

\section{Discussion}
\label{disc}

The properties of V32 are consistent with 
membership in the globular cluster NGC~6397.
The system is composed of two main-sequence 
stars with an age of about 12~Gyr.
The orbit shows a large eccentricity with $e=0.32$. 
Such an eccentricity is unexpected is since tidal 
forces should circularize the orbit
of the binary on a short time scale. The theoretical 
and observational aspects of the circularization 
of binary orbits for solar mass binaries 
are discussed in some detail in \citet{meibom}. 
They estimate, based on results of \citet{latham},
that halo binaries with periods shorter than 
$15.6^{+2.3}_{-3.2}$~d should have circular orbits.
For the old open clusters NGC~188 (age~$\approx$~7~Gyr) 
and M67 (age$~\approx$~4~Gyr) circularization has occured 
for periods shorter than 14.5~d and 12.1~d, respectively.

Very little is known about globular cluster binaries
with orbital periods longer than a few days. Several surveys for spectroscopic
binaries among cluster giants were conducted in 1980s and 1990s. They led to
the detection of some candidates but generally spectroscopic orbits of these 
stars remain undetermined \citep[see also references therein]{yan96}.
Four eclipsing binaries with periods $4<P<10$
days were detected in the central part of 47~Tuc in the HST/WFPC2 survey by 
\citet{albrow}
using photometric observations. All of these have presumed 
circular orbits although observations do not cover the full
orbital cycle for two  of the systems with the longest periods. 
Extensive ground base surveys of 47~Tuc 
\citep{weldrake04} and $\omega$~Cen \citep{kal04,weldrake07}
led to the detection of two eclipsing binaries with orbital periods exceeding
10 days. Follow up observations conducted by our group show that 47~Tuc-V69 and
$\omega$~Cen-V406 have eccentric orbits with $P=29$~d and $P=71$~d, respectively
(Thompson et al. 2008, in preparation).
All of the other cluster eclipsing binaries reported so far 
have orbital periods less than 10 days and show circular orbits.

The eccentric orbit of V32 can possibly be due to a relatively
recent dynamical interaction of the binary with other cluster star(s).
NGC~6397 has a ``collapsed'' core \citep{djorgovski} containing
a  population of about 20 X-ray sources including 9 candidate cataclysmic
variables, a millisecond pulsar and several candidate  BY~Dra-type close
binaries \citep{grindlay01}. \citet{kal06} 
reported the  detection of 9 eclipsing binaries and 6 candidates  for 
ellipsoidal binaries located in the central part of the cluster.
A dynamical interaction capable of transforming the circular orbit 
of V32 into a highly eccentric one would also likely
significantly disturb the systemic velocity of the binary. Our data show
that the radial velocity of V32 is not unusual  for an object
from the central part of the cluster. Unfortunately, the transverse 
velocity is V32 is unknown as the binary is not included in the recent 
proper motion survey of NGC~6397 conducted  by 
\citet{kalirai}.

As an alternative explanation of the eccentric orbit of V32 we
consider the possibility that the system is in fact a hierarchical triple.
The third component located on the outer orbit is capable of 
generating eccentricity in the inner binary. Extensive reviews of the 
evolution of the orbits of binary stars have been presented by \citet{egg06}
and \citet{mazeh}.
Numerical simulations of stellar clusters predict dynamical 
formation of  triple stars, especially in the presence of primordial binaries
\citep{mcmillan,heggie}.
Examples of known triple stars in globular clusters include the pulsar
PSR B1620-26 in M4 \citep{thorsett}
and possibly the ultracompact X-ray binary 4U~1820-303 
\citep{zdziarski}.
The hypothesis that V32 is a triple can be tested by spectroscopic
observations aimed at detection of variability of the 
systemic velocity of the putative inner binary. We see no 
systemic velocity residuals from the orbit give in Table 3 over the
2.2 year duration of our velocity observations.

Finally we  comment on possibility that the eccentric
orbit of V32 is related to its rather low mass ratio 
$q=M_{s}/M_{p}\approx 0.30$. According to \citet{mathieu}
the time scale of circularization of  a binary is related to $q$ by the relation:
$\tau_{circ}\propto q^{2/3}(1+q^{-1})^{5/3}$.
This corresponds to an increase of  $\tau_{circ}$ of a factor of 1.6 as
the mass ratio ranges from $q=1$ to $q=0.3$. 
For low values of $q$ the contribution of the 
secondary to the circularization process becomes negligible 
\citep{mazeh}. At the same time 
$\tau_{circ}$ is a strong function of the orbital period $P$, with
$\tau_{circ}\propto P^{16/3}$ \citep{zahn}.
The net result is that the circularization period will lengthen as the
mass ratio decreases from
for $q=1$ and $q=0.3$, especially for long period systems. 
It is appropriate to note at this point that 
that the sample of halo stars presented by \citet{latham}
consists entirely of single line binaries and hence objects with
mass ratios noticeably lower than unity.
More globular cluster binaries with period of the  
order of 10 days have to be detected and analyzed before 
V32 can be called a truly unusual system. The CASE group 
is in a process of collecting data which can shed more light
on the relation between eccentricity and orbital period for 
binaries in globular clusters.

\acknowledgments

JK and WK were supported by the grant 1~P03D~001~28  
from the Ministry of Science and Higher Education, Poland.
Research of JK is also supported by the Foundation for Polish
Science through the grant MISTRZ.
IBT was supported by NSF grant AST-0507325.
Support from the Natural Sciences and Engineering Council of Canada
to SMR is acknowledged with gratitude. It is a pleasure to thank
Willy Torres for sharing his code with us.

\clearpage


\begin{deluxetable}{lrrr}
\tablecolumns{4}
\tablewidth{0pt}
\tabletypesize{\small}
\tablecaption{Velocity Observations of V32 \label {rvdata}
\label{tab1}}
\tablehead{
\colhead{HJD}         &
\colhead{V$_{rad}$}     &
\colhead{$\sigma$ }       \\
\colhead{(- 2450000)} &
\colhead{km s$^{-1}$} &
\colhead{km s$^{-1}$} &
\colhead{}       
}
\startdata  
3521.85254 & 36.99 & 0.26\\
3581.71777 & 42.49 & 0.34\\
3582.69824 & 46.92 & 0.27\\
3584.66211 & 06.51 & 0.28\\
3875.88965 & 22.24 & 0.41\\
3876.71387 & 29.18 & 0.29\\
3877.72266 & 38.38 & 0.34\\
3891.82129 & -1.19 & 0.33\\
3898.77344 & 45.71 & 0.38\\
3935.65430 & 26.52 & 0.34\\
3938.69238 & 45.06 & 0.54\\
3989.54590 & 07.17 & 0.37\\
3991.58301 & 02.48 & 0.54\\
4258.67197 & 03.67 & 0.53\\
4259.73913 & 09.98 & 0.58\\
4329.66700 & 16.86 & 0.43\\
\enddata
\end{deluxetable}
\clearpage

\begin{deluxetable}{lrr}
\tablecolumns{2}
\tablewidth{0pt}
\tabletypesize{\normalsize}
\tablecaption{Orbital Parameters for V32\label{tab2}}
\tablehead{
\colhead{Parameter}       &
\colhead{Value}           &
\colhead{Value}
}

\startdata
$P$ (days)                    & 9.8783 $\pm$ 0.0011       & 9.8783 \tablenotemark{a} \\
$T_{1}$~(HJD-245 0000)        & 3899.832 $\pm$ 0.083      & 3899.889 \tablenotemark{a} \\
$\gamma~(km~s^{-1})$          & 20.67 $\pm$ 0.22          & 20.71 $\pm$ 0.18        \\
$K_{p}~(km~s^{-1})$           & 23.92 $\pm$ 0.30          & 24.00 $\pm$ 0.25        \\
$e$                           & 0.320 $\pm$ 0.011         & 0.322 $\pm$ 0.009      \\
$\omega$ (deg)                & 72.7 $\pm$3.0             & 74.74 $\pm$ 0.7        \\
$(O-C)~rms~(km~s^{-1})$      & 0.60                      & 0.62                   \\
Derived quantities:           &                           & \\
$f(M)~sin^{3}~i$~($M_{\odot}$)& 0.01192 $\pm$ 0.00046     &  0.01201 $\pm$ 0.00042 \\
$A_{p}sin(i)$~($R_{\odot}$)    & 3.079 $\pm$ 0.039         & 3.086 $\pm$ 0.036  \\
\enddata

\tablenotetext{a}{Fixed based on photometry.}
\end{deluxetable}

\clearpage

\begin{deluxetable}{lcc}
\tablecolumns{3}
\tablewidth{0pt}
\tabletypesize{\small}
\tablecaption{$V$-band Photometric Observations of V32
\label{tab3}}
\tablehead{
\colhead{HJD}         &
\colhead{$V$}     &
\colhead{$\sigma_V$ }       \\
\colhead{(- 2450000)} &
\colhead{} &
\colhead{}      
}
\startdata  
2765.8121 & 16.128 & 0.008 \\
2765.8209 & 16.134 & 0.008 \\
2765.8277 & 16.128 & 0.008 \\
2765.8371 & 16.126 & 0.008 \\
2765.8677 & 16.125 & 0.008 \\
2765.8894 & 16.129 & 0.008 \\
2765.8966 & 16.122 & 0.008 \\
2765.9061 & 16.129 & 0.008 \\
2765.9128 & 16.127 & 0.008 \\
2765.9294 & 16.131 & 0.008 \\
\enddata
\tablecomments{This Table is published in its entirety in the 
electronic addition of the {\it Astronomical Journal}. A portion 
here is shown for guidance regarding its form and content.}
\end{deluxetable}

\clearpage

\clearpage

\clearpage

\end{document}